\documentclass[mathleft
]{an}
\usepackage{graphicx}
\usepackage{times}
\usepackage{color}

\begin{document}

\Pagespan{1}{}
\Yearpublication{}%
\Yearsubmission{2012}%
\Month{}%
\Volume{}%
\Issue{}%

\title{A spectral differential approach to characterizing low-mass
  companions to late-type stars}

\author{N.M. Kostogryz\inst{1,2,3}\fnmsep\thanks{Corresponding author:
  \email{kosn@mao.kiev.ua}\newline}
   \and M. K\"{u}rster\inst{2}, T.M. Yakobchuk\inst{1},
   Y. Lyubchik\inst{1} \and M.K. Kuznetsov\inst{1}
          }

\titlerunning{Spectral differential companion characterization}
\authorrunning{N.M. Kostogryz et al.}
\institute{Main Astronomical Observatory of NAS of Ukraine 
(MAO NASU), Zabolotnoho 27, 03680, Kyiv, Ukraine 
\and Max Planck Institute for Astronomy,  K\"{o}nigstuhl 
17 D-69117 Heidelberg, Germany
\and Kiepenheuer-Institut f\"{u}r Sonnenphysik, Sch\"{o}neckstr.6, 
D-79104 Freiburg, Germany\\
}

\received{28 August 2012}
\accepted{}
\publonline{later}


\abstract{In this paper, we develop a spectral differential technique 
with which the dynamical mass of low-mass companions can be 
found. This method aims at discovering close companions to late-type 
stars by removing the stellar spectrum through a subtraction 
of spectra obtained at different orbital phases and discovering 
the companion spectrum in the difference spectrum in which the 
companion lines appear twice (positive and negative signal).
The resulting radial velocity difference of these two signals 
provides the true mass of the companion, if the orbital solution 
for the radial velocities of the primary is known. 
We select the CO line region in the K-band for our study, 
because it provides a favourable star-to-companion brightness 
ratio for our test case GJ1046, an M2V dwarf with a low-mass 
companion that most likely is a brown dwarf. Furthermore, 
these lines remain largely unblended in the difference spectrum 
so that the radial velocity amplitude of the companion can be 
measured directly.  Only if the companion rotates rapidly and 
has a small radial velocity due to a high mass, does blending 
occur for all lines so that our approach fails. 
We also consider activity of the host star, and show
that the companion difference flux can be expected to have 
larger variations than the residual signal from the active star
so that stellar activity does not inhibit the 
determination of the companion mass. In addition to 
determining the companion mass, we restore the single companion 
spectrum from the difference spectrum using singular value 
decomposition.}

\maketitle

\section{Introduction}
A large number of substellar companions to normal stars have 
been found to date, and substantial effort is devoted by the 
astronomical community to determine the characteristics of 
these objects. The most successful technique for the discovery 
of such objects is the radial velocity (RV) method.
Unfortunately, it leads to a rather limited knowledge of a 
detected companion. A precise determination of the mass of 
the companion object is not possible unless the data can be 
combined with observations obtained with other techniques 
(e.g. transits or astrometry) or constrained by the dynamical 
interaction effects occurring in multi-planet systems. The 
radial velocity method alone can only provide a minimum mass 
that, even though it is also the most probable mass, cannot 
by itself provide an unambiguous decision about the true 
nature of the companion object. Therefore, it is very important 
to develop new approaches to characterising low-mass companions 
to stars that provide a dynamical mass of the companion object. 

The present paper introduces a spectral differential method 
to the suite of techniques with which the dynamical mass and 
therefore the true nature of low-mass companions can be 
established. We develop our technique around the example of 
GJ1046b, which is a companion to the inactive old M2.5V star 
GJ1046 that was found by K\"{u}rster et al. (2008; see also 
Zechmeister et al. 2009) in their M dwarf RV program with the 
Ultraviolet and Visual Echelle Spectrograph (UVES) at the ESO 
VLT. This object has a minimum mass of $27$ 
Jupiter mas ($\mathrm{M}_\mathrm{Jup}$) 
that was derived from RV measurements. From a combination of 
the radial velocity measurements with Hipparcos astrometry 
(\cite{kuerster}) determined a confidence of 97$\%$ that the 
companion is a brown dwarf (i.e. a chance probability of just 
3$\%$ that it exceeds the stellar mass threshold). 
With a separation of 0.42 AU the companion is close to its host 
star and consequently it is located in the so called brown dwarf 
desert region. The orbital period is 168.8 d, the RV semi-amplitude 
is $1.83~\mathrm{kms}^{-1}$.  The orbit is moderately eccentric 
with an eccentricity of 0.28.

The brown dwarf desert is an observational range 
of orbits around a star in which brown dwarf companions are very 
rare (\cite{klahr, grether}). Its existence has been identified for typical 
star-companion separations of up to $\sim5$ AU. The origin of the 
brown dwarf desert is currently not fully understood and subject 
of a considerable amount of research (\cite{stamatellos}). 
What its existence seems to indicate is that two distinctive 
formation mechanisms are at work for planetary and stellar 
companions with relatively little overlap between the two. At wide 
separations ($>$1000 AU) no brown dwarf desert is observed. 
Originally defined for solar-type stars the brown dwarf 
desert is now also known to exist for early M dwarfs due to 
the small number of close-in brown dwarf companions to these stars. 
But as the mass ratios of binary systems with low-mass primaries 
tend towards unity, brown dwarf companions to late M dwarfs become 
more frequent (\cite{montagnier}). 

Around solar-type stars, few brown dwarf companion 
candidates in the separation regime up to a few AU are known. 
The first such candidate was HD 114762 (\cite{latham, cochran, mazeh}), 
followed by HD 168443c (\cite{marcy}), HD 202206b (\cite{udry}), 
HD 137510 (\cite{endl}), HD 191760 (\cite{jenkins}) and others. 
A transiting brown dwarf has also been discovered by the CoRoT
spacecraft. This object, CoRoT-Exo-3b, has a mass of 
$21.7~\mathrm{M}_\mathrm{Jup}$ and a radius of 1.01 Jupiter radii 
($\mathrm{R}_\mathrm{Jup}$), in a remarkably close orbit at 0.05 AU 
(\cite{deleuil}). The most recent brown dwarf companion in the 
brown dwarf desert range is TYC 1240-00945-1 (\cite{lee}). As will 
be pointed out in Section 3, the star-to-companion brightness 
contrast is an important parameter for our technique. For the 
mentioned objects this contrast is very large.

A more favourable brightness contrast can be found in systems 
comprised of a brown dwarf orbiting an M-type star in the brown 
dwarf desert range. Such objects can typically be found with RV 
measurements. There are several known candidates for probable 
brown dwarf companions to M-type stars, such as GJ1046 
(\cite{kuerster}), GJ595, GJ623, and GJ84 (\cite{nidever}). 

For this paper we chose GJ1046 as a sample to develop our 
technique, but certainly, we can apply this method to other 
brown dwarf companions to M-type stars.

\section{Synthetic spectra}

We computed synthetic spectra for M dwarfs and brown dwarfs 
using the WITA612 programme (\cite{pavlenko}). All calculations 
were carried out under the assumption of local thermodynamic 
equilibrium, hydrostatic equilibrium, and absence of sources 
and sinks of energy.  
The solar abundances reported by \cite{andgrev89} were used 
in the calculations. All details of the other input parameters
are described by \cite{pavlenkoetal}. 

A grid of synthetic spectra was computed from model structures 
of AMES-cond and AMES-dusty (\cite{allard01}) for the following 
range of parameters: temperatures of 1500~K, 2000~K, 2500~K and 
3500~K, $\mathrm{log}~g=5.0$, microturbulent velocity  
$2~\mathrm{kms}^{-1}$ and solar metallicity.

The atomic line list used for the computation of synthetic spectra 
is taken from the Vienna Atomic Line Database (VALD; Kupka et al. 1999). 
The main contributors of the molecular absorption around $\sim~2.3 \mu m$ 
are water and carbon oxyde. In our computations we used line lists BT2 
for H$_2$O by \cite{barber06} and CO by \cite{goorvitch}.

All theoretical spectra were computed with a wavelength
step of 0.01\AA~ and convolved with a Gaussian profile 
to match the spectral resolution of the instrument to be used for 
the observations. Rotational broadening of spectral lines is 
implemented by convolution with a rotational profile following 
\cite{gray}:

\begin{eqnarray}\nonumber
   G(\Delta\lambda) = (2(1-\epsilon)[1-(\Delta\lambda/\Delta\lambda_L)^2]^{1/2}+ 0.5\pi\epsilon[1- \\
 -(\Delta\lambda/\Delta\lambda_L)^2])/(\pi\Delta\lambda_L(1-\epsilon/3)) 
\end{eqnarray}

where $\Delta\lambda_L$=($\lambda$ vsin$i$)/c, $\epsilon$ is the 
limb darkening coefficient. Typically for our computations we have 
adopted a value of 0.6 which is sufficiently close to the K-band 
value for ultra-cool companions of 1500 K - 2500 K (\cite{claret}), 
and the dependence on this value is very small.

\section{Method of difference spectrum}

Our method aims at discovering companions to late-type stars by 
removing the stellar spectrum through subtraction of spectra 
obtained at different orbital phases and identifying the companion 
spectrum in the difference spectrum. Using near-infrared spectra 
of sufficient resolution one can attempt to search for absorption 
lines of the companion in the spectrum. In our approach we take one 
spectrum each near the maximum and the minimum of the radial velocity 
curve of the star, shift them in wavelength in such a way that the 
stellar line systems co-align and subtract them from each other thereby 
removing the stellar signal (not the photon noise, of course, that 
still has to be dealt with). What one is left with is a spectrum that 
consists of the difference of two spectra of the companion, but taken 
at different radial velocities. The companion lines will therefore appear 
twice in this difference spectrum, once as a positive signal and once as a 
negative signal. By determining the radial velocity difference of these 
two signals one can obtain the star-to-companion mass ratio and, if 
the stellar mass is known, one can determine the true mass of the companion 
and thereby its nature. 

The spectrum of the invisible companion will be detected when the 
star-to-companion brightness ratio (hereafter called "the contrast") of the 
two components of the system is sufficiently small. The smaller the contrast, 
the higher the probability to detect cool companions. Obviously, the 
best wavelength range for low-mass star observations is the near-infrared 
(NIR). \cite{knapp} provide the absolute magnitude - spectral type dependence 
in the J and K bands for objects from L0 to T9, from which we infer that 
in the J band the contrast is less favourable for spectral types L0-T0 
than in the K band while the contrast of late T dwarf companions relative to GJ 1046
is smaller in the J-band than in the K-band. 
Figure 1 shows the J and K band flux ratios 
of various field brown dwarfs as they would be observed 
at the distance of GJ 1046 with 
spectral types between L1 and T8 and corresponding temperatures (\cite{testi}). 
The data were taken from \cite{knapp}.   

\begin{figure*}
 \includegraphics[angle=270, width=1.0\textwidth]{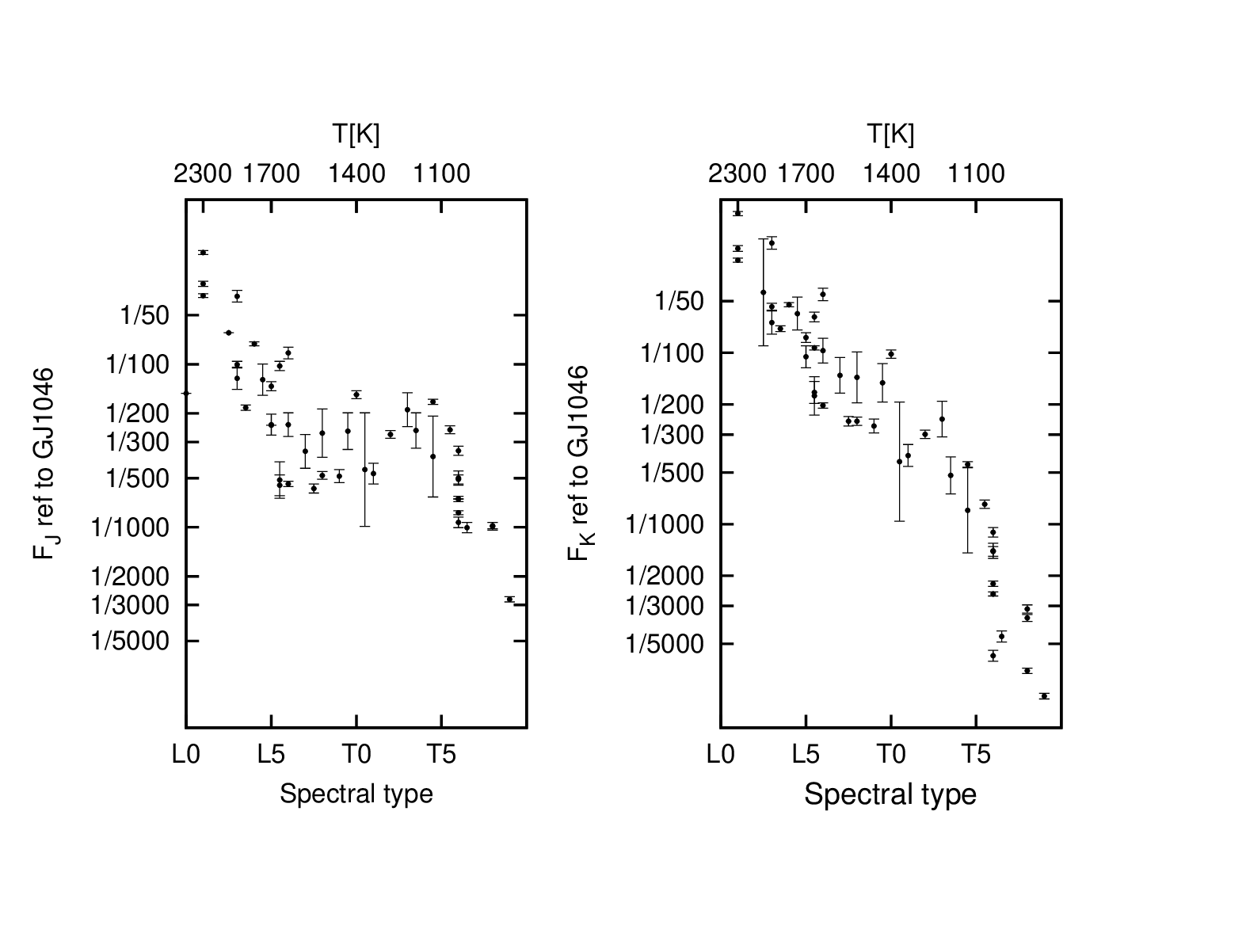}
 \caption{ The J and K band flux ratios relative to GJ 1046 
 of field brown dwarfs with spectral types from L0 to T9. 
 Temperature values (upper x-axis) that correspond to the different spectral 
 types were taken from Testi (2009).}
\label{fig: contrast}
\end{figure*}

To realize our approach, observations should be made at two 
epochs with different stellar radial velocities. In order to 
resolve the spectral features we plan to study we need an NIR 
spectrograph with high resolution of the order 50,000 - 100,000. 
The most suitable instrument for this technique is the cryogenic 
high-resolution infrared spectrograph (CRIRES) that is mounted 
at the Very Large Telescope of the European South Observatory. 

The absence of high-resolution spectroscopic observational 
data for GJ1046 leads us to simulate CRIRES observations in 
order to investigate the feasibility for the detection of the 
companion in the difference spectrum. For this purpose we take 
the synthetic spectra described in Section 2. As we do not know 
the spectral type of the companion, we take spectra with effective 
temperatures of 1500~K, 2000~K and 2500~K as samples to find out 
which of them can be detected. All these spectra were convolved 
with Gaussian instrumental profiles whose width can be chosen to 
match the resolution and spectral sampling of different instruments. 
Using cubic spline interpolation we map all these spectra to a sampling 
grid representing that of CRIRES. 

The selection of the spectral range is important for our approach.
Depending on the system, one needs to choose a wavelength range with 
small contrast and narrow lines. Since they are strong features, 
we have first considered the KI doublets in the J band but then we 
rejected them because their widths lead to blending problems (see below). 
The strong KI doublet lines in the J band become self-superimposed 
in the difference spectrum. Blending leads to an apparent increase of 
the wavelength shift and underestimation of the companion mass. 

Therefore, in our approach we switched to longer wavelengths (K band), 
which is preferable in order to do away with the blending problem and, 
besides that, having better contrast and signal-to-noise ratio. 

\begin{figure*}
 \includegraphics[angle=270, width=0.85\textwidth]{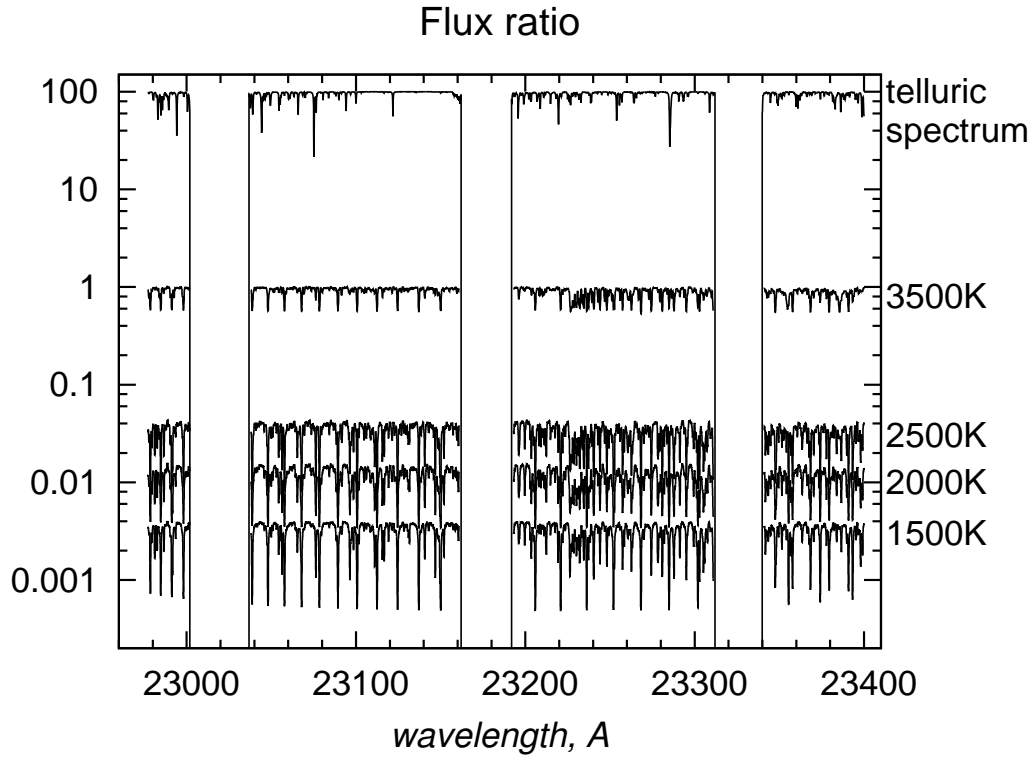}
 \caption{NIR spectra for 3500 K, 2500 K, 2000 K and 1500 K effective 
 temperature with their absolute flux corresponding to the mean contrast 
 for field brown dwarfs (cf. Fig. \ref{fig: contrast}). The upper spectrum 
is the transmission spectrum of the Earth's atmosphere taken from 
the CRIRES exposure time calculator 
($http://www.eso.org/observing/etc/bin/gen/form?INS.MODE=lwspectr+INS.NAME=CRIRES$). 
The three gaps in the wavelength regime near 2.302 $\mu m$, 
2.319 $\mu m$, and 2.332 $\mu m$ are due to physical gaps in the array 
of the four Aladdin detectors in CRIRES. Note also that due to vignetting 
the outer two detectors provide a shorter wavelength coverage than the inner two.}
 \label{fig:contrast1}
\end{figure*}

Also, the narrow spectral lines of the CO molecule are prominent 
throughout spectral types late-M to T in K band. We choose the CRIRES 
wavelength range order 24, standard settings, reference wavelength 
2.3252 $\mu m$ (see the CRIRES User's Manual, 
$http://www.eso.org/~rsiebenm/crires$) because it allows us to observe 
CO lines in the inner two detectors. Note that the outer two detectors 
of CRIRES provide a short wavelength coverage. Figures 2-5 are presented 
in this CRIRES wavelength range. The synthetic spectra of the host star 
and companion are presented in Figure \ref{fig:contrast1}. For the 
chosen spectra of effective temperatures 2500 K, 2000 K and 1500 K the 
absolute flux ratios in K band are 1/20, 1/60 and 1/250, respectively
(see Fig.\ref{fig: contrast}).

As is seen from Fig. \ref{fig:contrast1} the CO lines are contaminated 
by telluric absorption lines which must be divided out from the 
observational spectrum using synthetic telluric spectra (\cite{clough92, 
clough95, clough05, rothman, seifahrt}).

We assume that the orbital solution for the primary is well known 
from optical radial velocity data so that the RV of the primary can 
be predicted for both spectra. The uncertainties of the orbital solution of the primary are very 
small and do not affect the results derived with the technique 
presented here (see Section 4).
Provided that a perfect correction can be achieved for the stellar 
radial velocity difference between the two spectra to be subtracted 
from each other, which includes also a correction of the barycentric 
motion of the Earth, we can assume that the stellar spectrum cancels 
(see, however, Section 4). Another possible way to achieve the removal 
of the stellar spectrum would consist of aligning the two spectra via 
cross-correlation and then subtracting them from each other.  However, 
the stellar spectra are contaminated to an unknown amount by the 
companion signal which may affect the result of the cross-correlation 
and make this approach unreliable, so that we do not follow it.
For the purpose of our simulations it suffices then to subtract the 
companion flux $f(\lambda_i)$ from flux $f(\lambda_{i + \Delta i})$, 
which is shifted in wavelength by an amount $\Delta\lambda$ 
approximated by the nearest-integer shift in pixels $\Delta i$. 
The value of $\Delta\lambda$ can be calculated according to the 
Doppler formula as 

\begin{equation}
 \Delta\lambda = \frac{\Delta RV M_1}{c M_2} \lambda
\end{equation}

where 
$\Delta RV$ is the RV difference of the host star at the two 
different epochs and equal to twice the RV 
semi-amplitude of the star, 
$2 K_1$, when one observes at the extrema, 
$M_1$ is the host star mass,  
$M_2$ is the companion mass. 

Consequently, the difference spectrum is calculated using the equation:
\begin{equation}
 df(\lambda_i) = f(\lambda_i) - f(\lambda_{i + \Delta i})
\end{equation}

We generate Poisson distributed photon noise with a signal-to-noise 
ratio (SNR) equal to 500 per spectral pixel for each of the 
spectra from the different observational epochs. When taking the 
difference of the two spectra the photon noise is propagated accordingly.

Having obtained the input model difference spectrum (see below 
Fig.\ref {fig:companion1}) we can proceed to the procedure that 
simulates the companion mass determination. Practically, it consists 
of an accurate determination of the wavelength shift. In the absence 
of blending the spectral shift can be found as the simple positional 
difference between the positive and negative peaks averaged over all 
lines. 

\begin{figure*}
 \includegraphics[angle=270, width=0.95\textwidth]{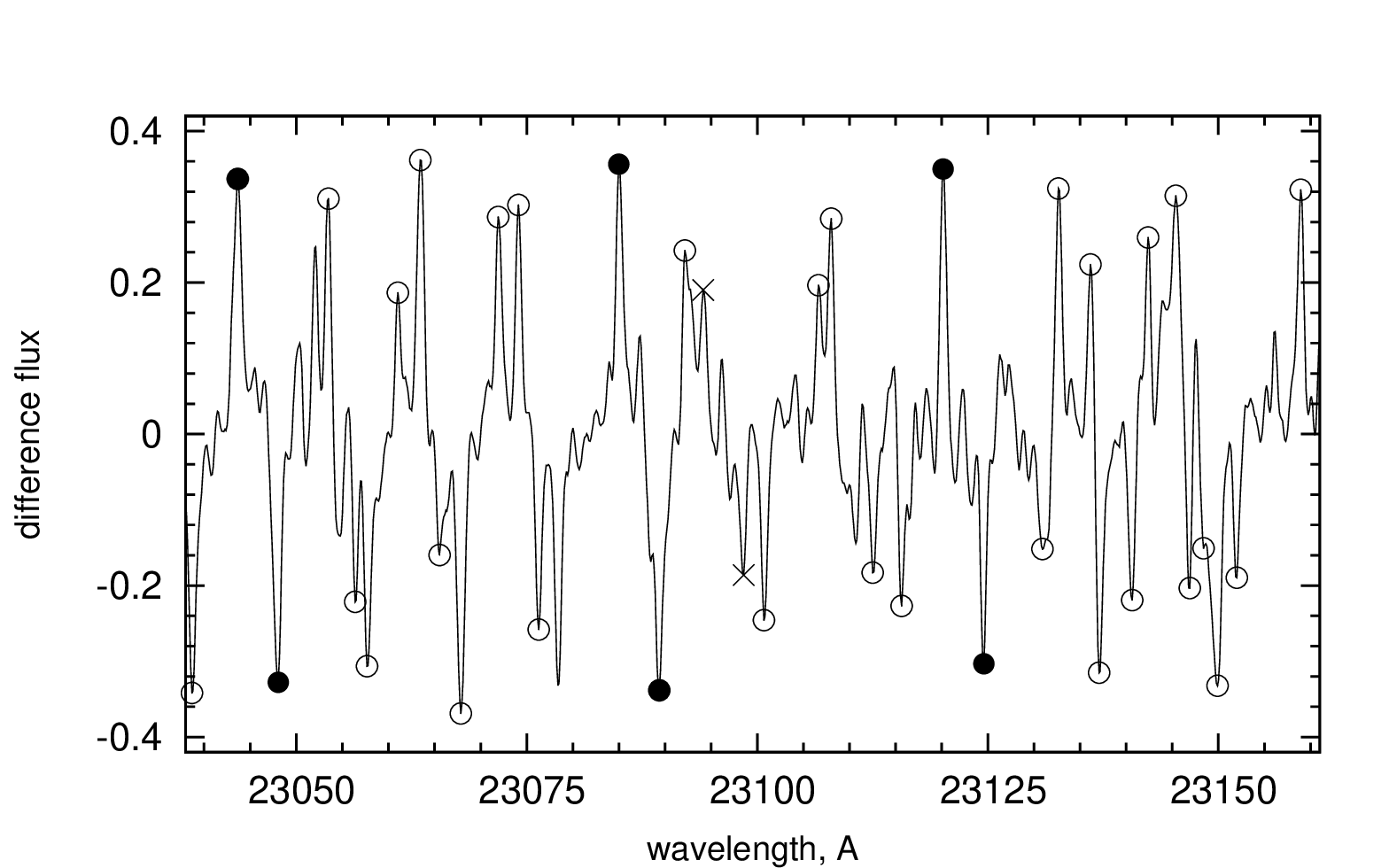}
 \caption{Difference spectrum for a non-rotating companion with the 
 minimum mass and 2500 K effective temperature. Shown is the spectral 
 range from 23030 \AA $~$to 23160 \AA $~$that corresponds to the wavelength range 
 of the second CRIRES detector.  Open circles depict all automatically 
 identified peaks that do not survive the filtering process (see Sect. 3).  
 Crosses mark automatically identified peaks that do not survive the visual 
 inspection.  Filled circles indicate peaks that survive to the end.
}
\label{fig:approach}
\end{figure*}

Overall, the procedure was performed as follows. First, we determine 
an initial guess shift value as the most frequent 
from visual inspection of all features.
Then we register all line peaks that exceed a selected positive or 
negative intensity threshold. 
Then we filter out unpaired peaks, paired 
peaks with intensity difference larger than the value adopted as initial 
guess and those with separation beyond the adopted shift range (adopted 
value $\pm$ 0.5). 
An additional check can be done visually to remove 
lines with bad or blended profiles (Fig.\ref{fig:approach}). 
Finally, having selected a suitable 
set of lines we can find the average spectral shift  
$\Delta i_{out}$, its errors and the 
corresponding companion mass value $M_{out}$ and error. 
Based on our model difference 
spectra we find that the presented procedure is successful enough to 
recover the input companion mass (see below Table 1).
 
\section{Host star M2.5V dwarf}

In this section we consider the impact of 
uncertainties related to the primary star on the 
results derived with our spectral differential technique.

\subsection{Misalignment due to wavelength calibration errors}

Apart from the blending issue, we consider that the largest source 
of systematic error is likely 
to come from errors in the wavelength calibration leading to imperfect 
alignment of the shifted stellar spectra before they get subtracted from 
each other. We therefore simulated the residual host star spectral features 
that would arise from this mismatch. As input spectrum for the M2.5V host 
star we use a synthetic spectrum for a temperature of 3500 K. 
  
We introduce shift values of 1.0, 0.5, 0.1, and 0.05 pixels before 
subtracting the spectrum from the so shifted version of itself. 
For the worth case of a non-transiting star (see below) Figure \ref{fig: host1} 
shows the residuals and indicates the photon noise level.  
As can be seen from Figure \ref{fig: host1}, the difference spectrum for the 
host star is beneath the noise level. While considerably higher for shift 
values of 1.0 and 0.5 pixels, the height of the strongest peaks are only at 
2.82 sigma for the 0.1 pixel shift value and at 1.39 sigma for 0.05 pixels.
Therefore, during the data reduction one should aim to achieve wavelength 
calibration errors less than 0.1 pix.  
Using the telluric lines present in the spectra \cite{brogi} obtained 
wavelength calibration errors of $0.15 - 0.20 \mathrm{kms^{-1}}$
for CRIRES in the same spectral region that we consider. As the pixel 
(px) width in CRIRES is equal to $1.5 \mathrm{kms^{-1}}$, the errors 
in the calibration are $0.1-0.13 \mathrm{px}$.
These values are the upper limit for our technique, therefore, we plan to use 
all possible options for the wavelength calibration to achieve smaller errors.
There are several other options for the wavelength calibration in ESO's standard 
calibration plan like ThAr spectrum or an $N_{2}O$ gas cell that can be used during spectra reduction. 

 \begin{figure*}
 \includegraphics[angle=270, width=0.85\textwidth]{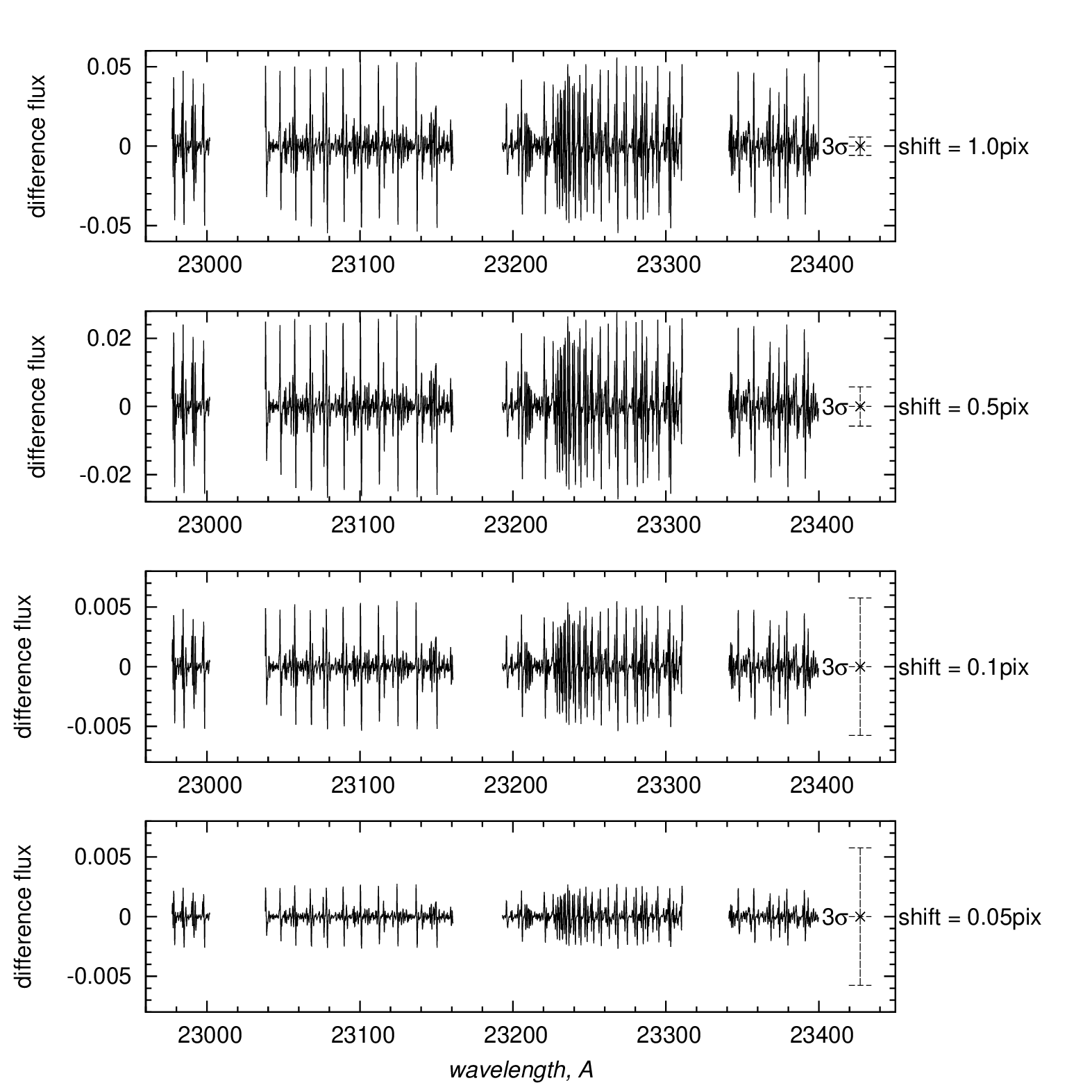}
 \caption{Difference spectra for a non-rotating host star with a 
wavelength calibration mismatch of 1.0, 0.5, 0.1 and 0.05 pixels. 
Note that for display purposes the photon noise has not been added 
to the signal, but is indicated by the 3-sigma error bars near the 
right end of the panels.}
 \label{fig: host1}
\end{figure*}
 
\subsection{Uncertainties in the orbital solution of the primary}
Uncertainties in the parameters of the orbital solution can lead to
systematic errors in the assumed radial velocity shift of the stellar
spectrum.  This can affect the precision with which the two stellar
spectra (taken at two different orbital phases) can be aligned before
they are subtracted from each other.

The true orbital phase may differ from the calculated one due to
the error in the orbital period and due to the error in the
periastron passage time.
While the latter is constant and known, the former is also known but
proportional to the number of orbital cycles elapsed between the time
of observation and the epoch at which the ephemeris was determined.

According to K\"{u}rster et al. (2008) the orbital period for 
GJ1046 is $168.848 \pm 0.030 d$ and the
time of periastron $Tp = BJD 2 453 225.78 \pm 0.32 (d)$. 
The ephemeris were determined at epoch 2006.0 while 
we are planning for observations for epoch 2014.0. The difference of the 
epochs is equal to 8 yr, i.e. 2922 d, i.e. 17.31 orbital cycles.
Considering this difference the accumulated phase error is 0.519 d. 
The combined uncertainty including the accumulated phase error and
the error of the periastron time the maximum uncertainty is 0.61 d.

The maximum RV change (occurring on the descending branch of the RV
curve where the RV is equal to the systemic RV which is
actually an uninteresting phase for our technique) is 
$-0.125 \mathrm{kms^{-1}d^{-1}}$ (as can be inferred from 
Fig. 1 in \cite{kuerster}), 
i.e. $-0.0801 \mathrm{pxd^{-1}}$ as the pixel width for CRIRES is $1.56 \mathrm{kms}^{-1}$.
The uncertainty in the RV due to the error in phase is, therefore, less than 
$0.077 \mathrm{kms}^{-1}$, i.e. $< 0.049\mathrm{px}$ 
everywhere on the RV curve, and much less near the extremal values of the curve.    
For the difference of two observations this uncertainty will in most
realistic observational scenarios be reduced further. 

However, it can be increased when observing at the descending and ascending     
branches. So, the worst case is when observations are carried out near 
the systemic RV, where in the ascending
branch a velocity up to $0.0399 \mathrm{kms^{-1}d^{-1}}$, i.e. $0.0256 \mathrm{pxd^{-1}}$ can be reached
leading to an uncertainty of the phase of $0.0207 \mathrm{kms^{-1}}$, i.e. $0.0133 \mathrm{px}$
(the uncertainty in the constant time of periastron passage is not
included again). In this worst case the combined uncertainty is then
$< 0.080 \mathrm{kms^{-1}}$ or $< 0.051 \mathrm{px}$. As it is seen from Fig.\ref{fig: host1} even in the 
worst case the uncertainties of the primary's orbital solution can be neglected
in our technique. But the observations are planned at phases where the RV changes
are expected to be much slower so that the uncertainties will also be much smaller.

\subsection{Activity of the host star}
Another source of systematic error in our approach could be the 
activity of the star, i.e. star spots. As we need to observe at different 
observational epochs our approach may be susceptible to changes 
in the stellar surface temperature distribution arising from appearing 
and disappearing star spots. The influence of stellar activity on the 
determination of the companion mass with our difference spectrum technique 
is the subject of a set of simulations presented in the following. 

It is known from the Sun that magnetic activity can cause
spots which are significantly cooler than the quiet photosphere. 
However, for very active stars the tempreture difference of spots and photosphere 
remains a rather unknown parameter. According to \cite{solanki} the 
coolest temperature for sunspots is about 2000K lower than for the photosphere. 
For active G- and K-type stars the spots can cover between 10\% and 50\% of 
the stellar surface and have temperatures up to 
1500K below photosphere (\cite{oneal01, oneal04, reiners}). 

From available observations, no information is available on 
spot distributions in M dwarfs, i.e. in stars with 
atmospheres that differ substantially from those of Sun-like stars.
Therefore, we make the following type of simulations.
We assume a single spot that is a function of 
\begin{itemize}
 \item spot surface filling factor;
 \item spot-to-photosphere temperature ratio;
 \item the instantaneous radial velocity of the spot  on the rotating star.
\end{itemize}

\begin{figure}
 \includegraphics[angle=0, width=0.45\textwidth]{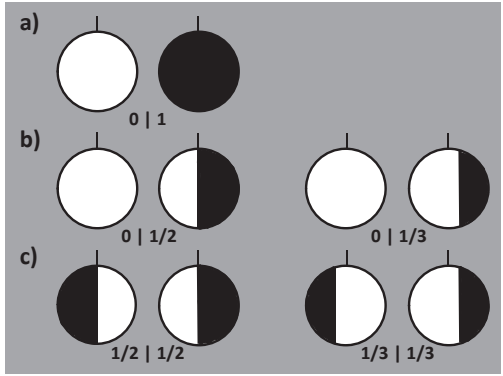}
 \caption{Illustration of spot distributions on the host star
that are considered in our simulations. In case (a) the spectra observed at different two epochs will maximally differ from each other. In cases (b) and (c) there will also be RV shifts between tha pairs of spectra.}
 \label{fig: host2}
\end{figure}

In attempt to study worst case conditions we choose the spot geometries shown in Fig.\ref{fig: host2}. In the first example the star is spot free at one of the two observational epochs, whereas it is fully covered by the spot at other epoch (a). In this case there is a maximum diference between the two spectra of the star.In the two examples labelled (b) the spot filling factors of the visible hemisphere are $1/2$ and $1/3$, respectively. Assuming that the star rotatescounterclockwise when viewed fromabove, then the spot is completely placed on the receding hemisphere and covers the location of the largest possible RV values in order to maximize RV shift effects between the spectra taken at the two epochs. In the two examples labelled (c) RV shift effects are increased even further due to the fact that the spots now appear at both epochs, but at mirror-symmetric locations.

Reliable spot temperature values for early M dwarfs are not available. In order to study worst case effects we therefore choose a spot temperature that is 1000 K below the photospheric temperature (\cite{terndrup, scholz}, see also last paragraph of this section). The combined stellar spectrum is then composed by the spectrum of the unspotted photosphere and the spectrum of the considering the areas covered and the RV shifts.

\begin{figure*}
 \includegraphics[angle=270, width=0.85\textwidth]{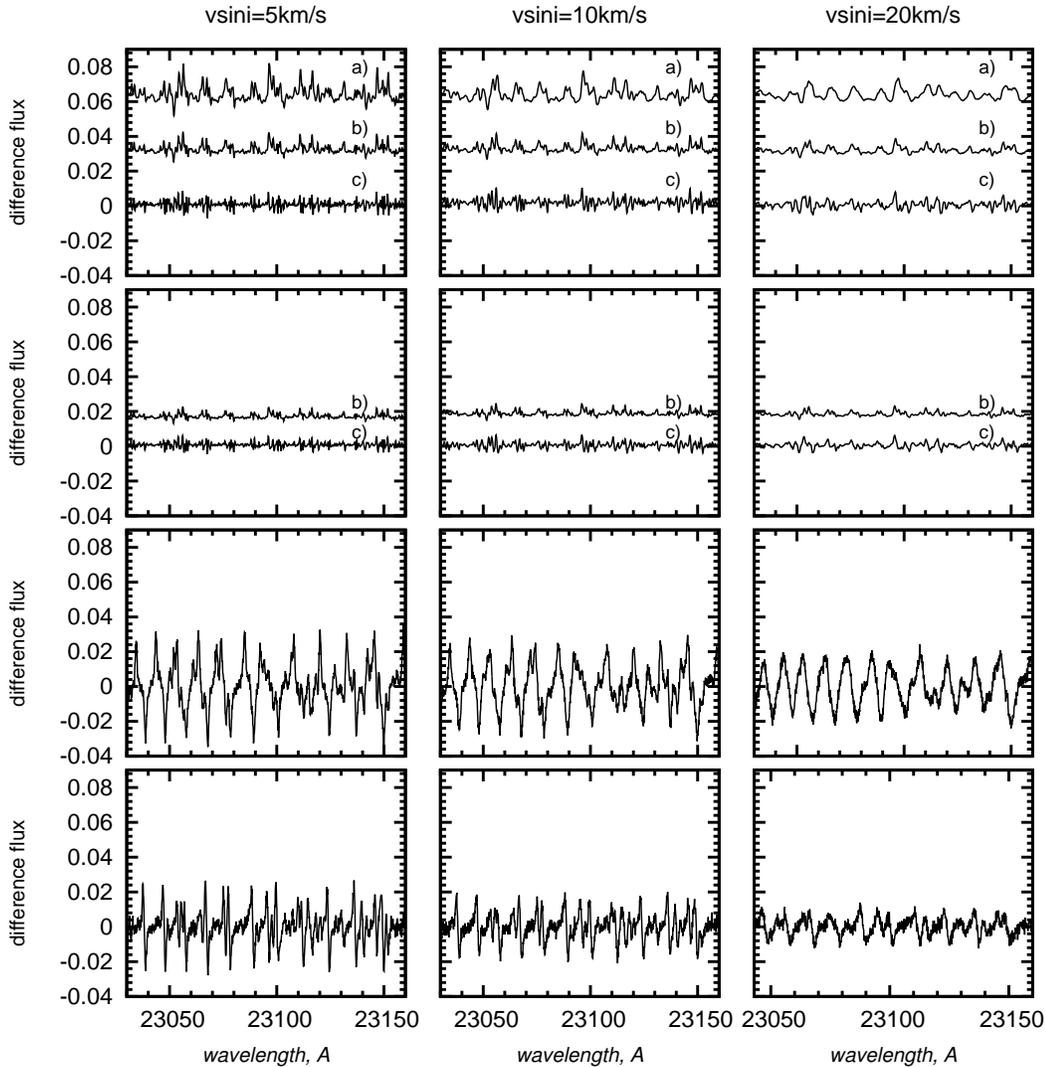}
 \caption{ Simulations of the residual signal from the active host star (upper two rows of panels)
 and the difference flux of the companion (lower two rows of panels) with an effective temperature of 1500K 
 for different rotational velocities.
 The first row corresponds to the first column of Fig.\ref{fig: host2}.
 The second row presents the results of the simulation when the filling factor is equal to $1/3$
 (second column of Fig.\ref{fig: host2}). The labels $"a)", "b)", "c)"$ to the geometry shown in Fig.\ref{fig: host2}. The next two 
 rows show the difference spectra of the companion with the minimum and
 the maximum masses, respectively. The signal in the companion difference 
 spectrum exceeds that produced by the activity effects in the host star 
 in almost all cases.}
 \label{fig: host3}
\end{figure*}

Considering the rotation of the star, we have to shift 
the cooler spectrum relative to the hotter one in wavelength according 
to the assumed velocity that results from stellar rotation. Convolving
the cooler spectrum with that part of the rotational profile (Eg.1) that 
corresponds to the spotted area, and the hotter 
spectrum with that part of the rotational profile that corresponds to 
the unspotted area of the star, 
adding these spectra and convolving the result with the instrumental 
profile of CRIRES, we obtain
the spectrum of the object during the individual epoch. The same procedure 
is applied to the second observational epoch. Then, taking a difference 
between the spectra from two epochs, we get the residual spectrum for 
the active star 
(Fig.\ref{fig: host3}). In Fig.\ref{fig: host3} the first row is 
plotted for visible hemisphere spot
filling factors of $1$ and $1/2$ and for the cases depicted in the 
first column of Fig.\ref{fig: host2}; the second row is the residual 
spectrum of the active host star corresponding to the second column 
in Fig.\ref{fig: host2}; in 
the third and forth rows we show the difference flux of the companion with 
effective temperature 1500K and for the minimum and maximum
masses of the companion, respectively. The three columns of the 
Fig.\ref{fig: host3} are for different rotational velocities of the star 
(first two rows) and the companion (last two rows), respectively. 
As star and companion do not necessarily have the same rotational velocity, 
one can compare the residual stellar and difference companion spectra for different 
rotational velocities. Fig.\ref{fig: host3} shows that in extreme cases, such as those we are studying, differences in the overall spot coverage between the two observational epochs can be more inportant than RV shifts.  
For the given spot temperature the large spot coverage difference corresponding to configuration (a) in Fig. \ref{fig: host2}, leads to stronger variations in the difference spectrum than RV shifts between spot and photosphere (configurations b and c), because latter can only be produced by a partial spot coverage.
As is seen from Fig. \ref{fig: host3} we can neglect the 
activity of the star except in the most extreme
case when we have a slowly rotating and a spotless star during the first epoch and completely spotted star during the next observational epoch (Fig. \ref{fig: host3}, top left panel, curve labelled $"a)"$), and 
fast rotation and maximum mass for the companion 
(Fig. \ref{fig: host3},  bottom right panel).

\begin{figure}
 \includegraphics[angle=270, width=0.45\textwidth]{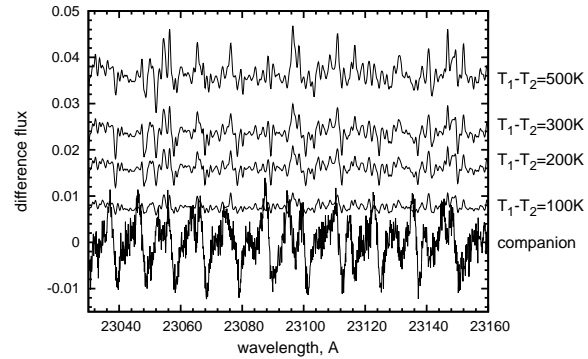}
 \caption{Simulation of the residual signal of the active star that has small 
 rotational velocity ($v\sin i = 5 \mathrm{kms}^{-1}$) and is very active (the geometry of the spot corresponds to the case $"a)"$ Fig. \ref{fig: host2})(thin lines) compared to difference flux of the companion with the maximum mass
 and rotational velocity of $20~\mathrm{kms}^{-1}$ (thick line).
The labels point at the temperature differences of the host star between the two observational epochs.
 }
 \label{fig: host4}
\end{figure}

However, the spot-to-photosphere brightness ratio is smaller in the NIR (\cite{terndrup, scholz}). For this worst case, i.e. maximum variations in the host star difference spectrum and minimum variations for the companion difference spectrum, we now study the effect of reducing the temperature difference between spot and photosphere to values in the range of 100 K to 500K. We simulate the residuals from the extremely active host star
(Fig. \ref{fig: host2}, a) that rotates
with $v\sin i = 5 \mathrm{kms}^{-1}$ and the difference flux of the fast rotating
($v\sin i = 20 \mathrm{kms}^{-1}$) companion that has maximum mass (Fig.\ref{fig: host4}).
In Fig.\ref{fig: host4} the thick solid line corresponds to the companion difference spectrum and the thin solid lines depict the residual signal from the host star for the considered temperature differences.
So, when the temperature difference of the star between the two epochs is 500 K, the activity could still be a problem for our technique, but it should be mentioned here that this case is quite unrealistic. If the star is so active that half of its surface is covered by a cool spot, it should rotate very fast which lead to broadening of its spectral lines and will result in a decreasing of the variations in the residual spectrum (see Fig. \ref{fig: host3}). And moreover, even for such an extreme case, the activity of the host star is not an issue any more for our technique when the temperature difference is smaller than 500 K. As was shown by Tendrup et al. (1999), for a star with a mass similar to that of GJ1046 the temperature contrast in the NIR is about 6\%, i.e. 200 K but the spot-filling factor is only 13\% that is smaller than in our simulation. So, the activity of the star should not be an issue in our technique.  

\section{Companion brown dwarf}

By planning for observations of GJ1046 near the minimum and maximum of its RV curve and 
then subtracting the obtained spectra from each other we aim at the discovery of the  
CO lines of the brown dwarf candidate companion to the M2.5V host star. 

\begin{figure*}
 \includegraphics[angle=270, width=0.75\textwidth]{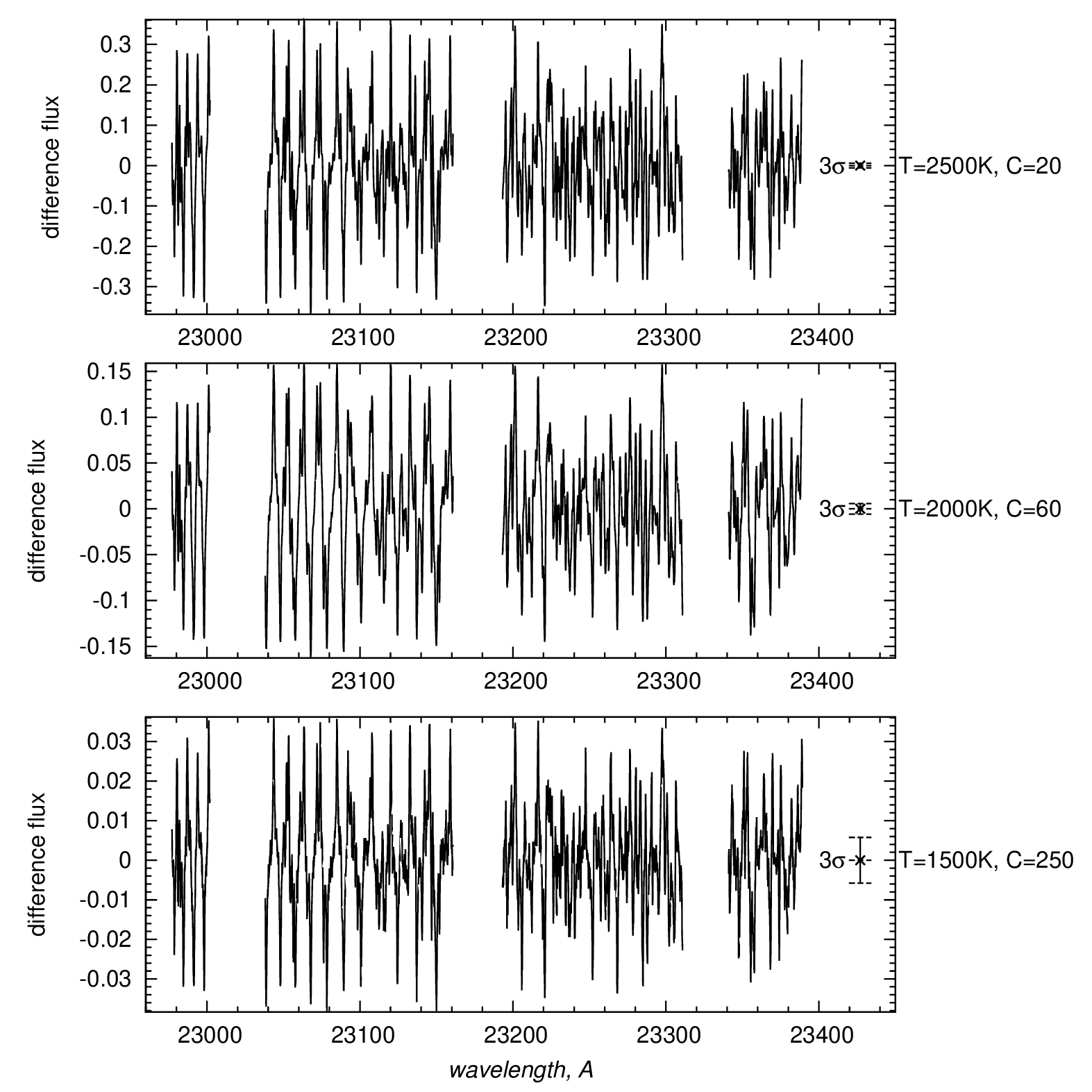}
 \caption{Simulated difference spectra obtained from observations at companion's RV maximum 
and minimum. The stellar spectrum cancels. Three panels for effective temperatures 2500 K, 
2000 K and 1500K, and contrasts C=20, 60, 250, respectively, for the minimum possible companion 
mass $m_{2, min} = 26.85 M_{Jup}$. The solid line is the companion difference flux with photon noise added,
the crosses together with the 3-sigma error bars depict just the photon noise. }
\label{fig:companion1}
\end{figure*}

Figure \ref{fig:companion1} presents simulations of difference spectra taken at maximum amd 
minimum companion RV, previously shifted to correct for stellar orbital and Earth's 
barycentric motions thereby making the stellar spectrum cancel. 

In the difference spectrum the most interesting structures are the combined absorption/pseudoemmision 
features produced by each of the absorption lines resulting from the wavelength shift between 
the two spectra of the companion. The observed signature at the CO line positions is stronger for brighter companions.
Note that even though the host star spectrum cancels, we are still left with the combined photon noise from 
the individual stellar spectra.

Experience with infrared detectors shows that very high signal-to-noise ratios (SNR) are difficult to achieve, because of several effects. 
The first such effect is the required high quality of flat-fielding in order to calibrate the strong pixel-to-pixel 
variations that these detectors have. The flat-fields provided by the CRIRES calibration plan for our selected 
wavelength setting possess an SNR ranging from about 1000 - 1400 per spectral bin (see $http://www.eso.org/observing/dfo/quality/
ALL/ref_fra\\mes/CRIRES/flat_sn/$). A second effect is detector non-linearities that effectively reduce the achievable
SNR.  They are a concern for the CRIRES Aladdin III detectors.
Additionally, the quality with which the telluric lines can be removed or modelled
will also influence the achievable SNR.
These considerations lead us to assume (perhaps somewhat optimistically) that the
maximum achievable S/N ratio with CRIRES will be about 500 per spectral bin, a 
value comparable to that attained also in other work, e.g. (\cite{rodler})
who obtained an SNR of 300 at the same wavelength setting also for high-precision work.

Taking a signal-to-noise ratio of 500, we determine the difference spectrum 
for the three cases of the companion with effective temperatures  2500K, 2000K, 1500K (Fig. \ref{fig:companion1} shows this
for the minimum mass but the result for maximum mass is very similar). As is seen 
from Figure \ref{fig:companion1}, the signal to noise ratio is high even for 
the object with effective temperature 1500K. Therefore, we can confirm that it is feasible to 
detect this difference companion spectrum in the planned observations with CRIRES. 
Moreover, the object with a temperature of 1500K is not the lower limit for detecting the companion 
so that cooler object can also be detected in the difference spectrum. 

Once we have obtained the difference spectrum, we will try to find the mass of the companion. 
For this purpose it is important to consider the broadening of the spectral lines due to 
rotation of the companion.  
The faster the rotation of the companion, the broader the CO lines will be 
and therefore more blended lines will appear in the difference spectrum.
All our synthetic spectra were convolved with a rotational profile corresponding to projected rotational velocities of 
$2~\mathrm{kms}^{-1}$, $5~\mathrm{kms}^{-1}$, $10~\mathrm{kms}^{-1}$, $15~\mathrm{kms}^{-1}$ and 
$20~\mathrm{kms}^{-1}$ to study how rotation of the companion influences 
the determination of the companion mass. 
The input values of the wavelength shifts and masses were taken from the difference spectrum as described in section 3 last paragraph. 
Then, we are looking for
similar ($\Delta i_{in} \pm 0.5$) shifts between the pseudoemission and absorption in the difference spectrum.
Because of the photon noise in the difference spectrum the obtained wavelength shifts may not correspond to the true ones.
Therefore, after identifying the correct features in the difference spectrum we fit double Gaussians to the positive and 
negative peaks of the difference spectrum features using the Levenberg-Marquardt 
method from Numerical Recipes (\cite{press, marquardt}). 

\begin{table*}
 \centering
 \begin{minipage}{165mm}
  \caption{Determined companion masses for different rotational velocities of the companion: $2~\mathrm{kms}^{-1}$, $5~\mathrm{kms}^{-1}$, 
$10~\mathrm{kms}^{-1}$, $15~\mathrm{kms}^{-1}$ and $20~\mathrm{kms}^{-1}$. Input shift value in pixels $\Delta i_{in}$ 
were taken from the difference spectra.}
\begin{tabular}{|c|c|c|c|c|c|c|c|c|}
 \hline    
  $v$  & $\Delta i_{in}$ & $M_{in}$  &  & $\Delta i_{out}$ & & & $M_{out}(M_{Jup})$& \\ \cline{4-9}
  $\mathrm{sin}i$& & $(M_{Jup})$ & 2500K & 2000K & 1500K & 2500K & 2000K & 1500K \\  
 \hline
 
 2  & 4.35 & 26.85 & 4.345$\pm$0.028 & 4.331$\pm$0.041 & 4.323$\pm$0.029 & 27.00$\pm$0.17 & 27.09$\pm$0.26 & 27.13$\pm$0.18\\
 5  & 4.35 & 26.85 & 4.327$\pm$0.035 & 4.355$\pm$0.052 & 4.332$\pm$0.022 & 27.11$\pm$0.22 & 26.93$\pm$0.32 & 27.08$\pm$0.14\\
 10 & 4.35 & 26.85 & 4.345$\pm$0.045 & 4.410$\pm$0.081 & 4.403$\pm$0.045 & 27.00$\pm$0.28 & 26.60$\pm$0.49 & 26.65$\pm$0.27\\
 15 & 4.35 & 26.85 & 4.295$\pm$0.061 & 4.435$\pm$0.098 & 4.370$\pm$0.067 & 26.80$\pm$0.35 & 26.45$\pm$0.58 & 26.85$\pm$0.41\\
 20 & 4.35 & 26.85 & 4.359$\pm$0.095 & 4.399$\pm$0.077 & 4.383$\pm$0.039 & 26.91$\pm$0.58 & 26.67$\pm$0.47 & 26.76$\pm$0.24\\
 \hline
 2  & 1.05 & 111.74 & 1.072$\pm$0.014 & 1.083$\pm$0.015 & 1.077$\pm$0.014 & 109.4$\pm$1.4 & 108.3$\pm$1.6 & 108.9$\pm$1.4 \\
 5  & 1.05 & 111.74 & 1.089$\pm$0.013 & 1.097$\pm$0.040 & 1.067$\pm$0.018 & 107.7$\pm$1.3 & 107.0$\pm$4.0 & 110.0$\pm$1.9 \\
 10 & 1.35 & 86.91  & 1.369$\pm$0.028 & 1.399$\pm$0.022 & 1.333$\pm$0.037 & 85.7$\pm$1.8  & 83.9$\pm$1.3  & 88.0$\pm$2.5 \\
 15 & 1.75 & 67.04  & 1.819$\pm$0.025 & 1.778$\pm$0.016 & 1.644$\pm$0.043 & 64.49$\pm$0.89& 65.98$\pm$0.60& 71.3$\pm$1.9 \\
 20 & 2.15 & 54.57  & 2.152$\pm$0.057 & 2.088$\pm$0.089 & 2.091$\pm$0.039 & 54.5$\pm$1.4  & 56.2$\pm$2.4  & 56.1$\pm$1.1 \\
 
\hline
\end{tabular}
\end{minipage}
\end{table*}

\begin{figure*}
 \includegraphics[angle=0, width=0.85\textwidth]{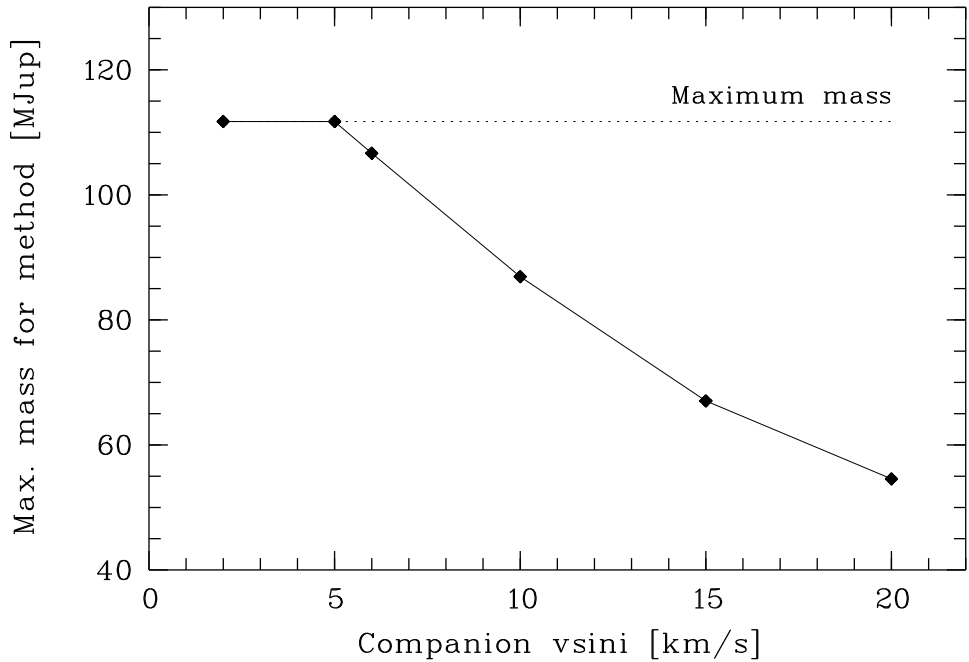}
 \caption{Maximum value of the companion mass for which our
difference spectrum approach yields a correct mass determination as a function
of the $v~\mathrm{sin}i$ of the companion (solid line with diamonds). The dotted line depicts
the correct maximum mass value. Due to line blending the determination of
masses in the regime between the two lines will yield values that are systematically
too small.}
 \label{fig: maxmass}
\end{figure*}

Table 1 shows the determined RV shifts and masses from the difference companion spectrum with various rotational velocities.  
For a companion with the minimum possible mass no blending effect occurs in the difference spectra 
which results in a correct mass determination (Table 1) for all adopted rotational velocities. 
But, for the maximum companion mass it is only possible for rotational velocities of 
 $2~\mathrm{kms}^{-1}$ and $5~\mathrm{kms}^{-1}$, to find the correct wavelength 
shift and therefore the 
mass of the companion (all obtained masses are within $\sim3~\sigma$ of the true mass). For higher rotational velocities, 
the obtained masses are wrong because of blending of the CO-lines in the difference spectrum 
(see Fig. \ref{fig: maxmass}).

\section{Companion spectrum reconstruction} 

To reconstruct the (single) companion spectrum from the difference spectrum, we apply the 
method of singular value decomposition using the algorithm by \cite{press} on difference spectrum. 
Using the shift value taken from Table 1, we reconstruct the 
single companion spectrum from the difference spectrum. 
For each pixels we rewrite the system of equations (Eq.2) as the matrix equation:

\begin{equation}
\left[ \begin{array}{c}
 df_1 \\ df_2 \\ \vdots \\ df_N 
\end{array} 
\right]  =  \left[ 
\begin{array}{cccc}
A_{11} & A_{12} & \cdots & A_{1M} \\ 
A_{21} & A_{22} & \cdots & A_{2M} \\
& \cdots & \\
A_{N1} & A_{N2} & \cdots & A_{NM} 
\end{array} 
\right] \times \left[ \begin{array}{c} 
f_1 \\ f_2 \\ \vdots \\ f_N 
\end{array} \right]
\end{equation}

where $N$ is the number of rows and $M$ is the number of columns of the matrix. Since $M > N$ the problem is ill-posed, i.e. underdetermined. 
The coeficients $A_{ij}$ are equal to 1 for $i=j$ and equal to -1 for $i=j+\Delta i$. In all 
other cases $A_{ij} = 0$. $\Delta i$ is the number of pixels corresponding to the wavelength shift $\Delta \lambda$. 

\begin{figure*}
 \includegraphics[angle=270,width=0.95\textwidth]{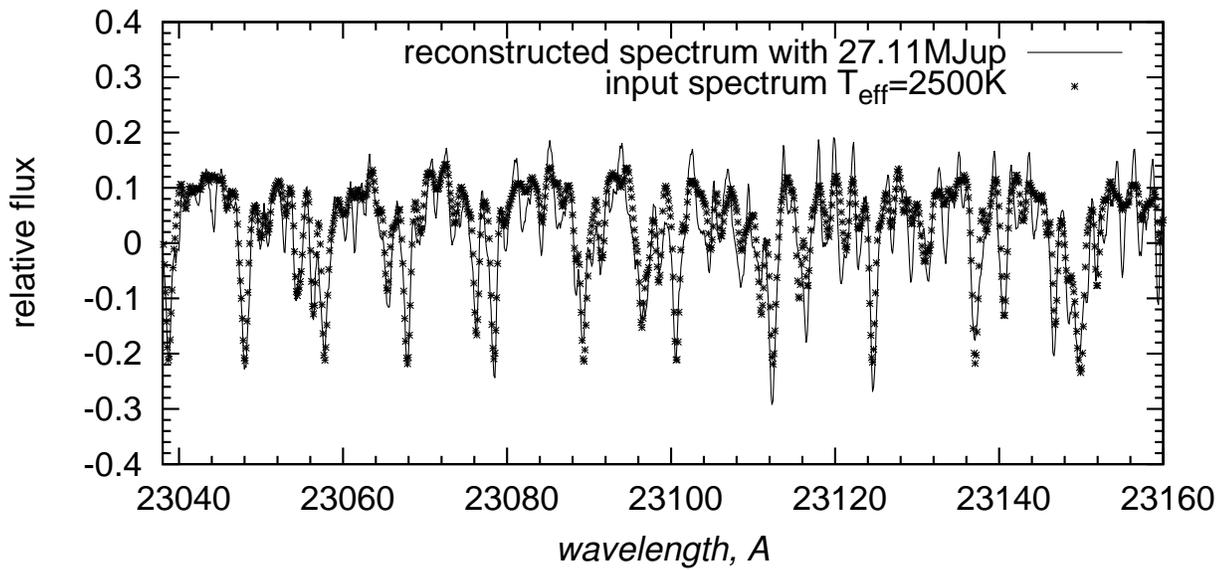}
 \caption{Companion spectrum (solid line) reconstructed from difference spectrum for a companion with 
a mass of $27.11 M_{Jup}$. The crosses show the input spectrum of the companion with the same 
signal-to-noise ratio.  
}
\label{fig:rec}
\end{figure*}

The reconstructed companion spectrum is shown in Fig.\ref{fig:rec}.  Note that for display purposes we show it only for the wavelength region 
covered by the second CRIRES detector in the wavelength setting selected by us.
We reconstructed this companion spectrum for the case of a companion mass of $27.11 M_{Jup}$, 
effective temperature of 2500 K, and rotational velocity of $2~\mathrm{kms}^{-1}$ (cf. Table 1).
As is seen from Fig.\ref{fig:rec}, the main features are reconstructed, but that there are also
some discrepancies in the finer details.

Knowing the spectra of the companion several of its parameters can be obtained, 
such as spectral type of the companion, effective temperature and age. 
  
\section{Discussion}
 In recent years a suite of differential observing techniques has
  been developed which have proven to be very useful tools to
  discover or characterize sub-stellar companions to stars.  An
  example is spectral differential imaging (\cite{smith, marois00, lenzen}) 
  which employs images taken
  through different filters in which the companion brightness
  differs due to a strong spectral feature, but in which the primary
  star brightness is unchanged.  It is essentially by taking the
  difference of these images that the bright star can be removed,
  and the companion is uncovered.  Another example is angular
  differential imaging (\cite{marois06, lafreniere, thalmann}) 
  which exploits field rotation in alt-az-mounted
  telescopes with deactivated derotator (so-called "pupil-tracking
  mode") to obtain a high fidelity removal of the primary without
  destruction of the companion signal.

  Quite recently, both \cite{rodler} and \cite{brogi} have exploited the CO line region
  in the K-band that we also use in our study, discovered the spectral
  signal of the Hot Jupiter companion to $\tau $~Boo, and determinedz
  its mass.  Due to the short orbital period of this system observations
  from a range of orbital phases were combined. \cite{rodler}
  used a differential approach for the removal of the stellar spectrum
  by co-adding all observations into a template that was then subtracted
  from the individual spectra (for earlier applications of this type
  of approach see, e.g., Cameron et al. 1999 or Rodler et al. 2010).  \cite{brogi} managed to
  ignore the weak stellar lines of the F7V star in their
  cross-correlation approach.

  In the present paper we provide a differential technique which
  also employs spectroscopy and leads to the determination of the
  RV amplitude and the mass of low-mass and sub-stellar companions to
  late-type stars with existing orbital solutions for the primary star.
  Methodologically, our approach bears some resemblance to that taken
  by \cite{rodler}.  However, it differs in that we restrict
  ourselves to the phases of the extrema of the RV curve which is
  easier to do for objects with longer periods such as GJ1046.
  Another difference is the application to M dwarfs which have their
  own strong absorption line system whose removal is more demanding
  than the treatment of the spectra of earlier-type stars.

  While it is the primary goal of our approach to measure the RV
  amplitude and hence the mass of the companion the fact that we also
  can approximately reconstruct the companion spectrum is a useful
  by-product for further characterization of the companion.  As can
  be seen from Fig. \ref{fig:rec} the reconstruction is not perfect which is a
  direct consequence of the fact that we employ only two orbital phases.

  In the literature several techniques to uncover the spectra of the
  companion object in binary star systems have been presented.
  \cite{simon} presented a technique for the disentangling
  of composite spectra requiring a series of these spectra taken at
  several orbital phases and, as in our approach, employing singular value
  decomposition.  They applied their method to an O-type binary with
  components of similar brightness.  \cite{hadrava} show how to decompose
  even the more complex composite spectra of multiple systems in Fourier
  space.  In other work template or model spectra are fitted to remove the
  primary spectrum and uncover the companion spectrum; see e.g. \cite{griffin} 
  who in this case study an F-type companion to a K giant.

  The case we present in this paper differs from these literature
  applications in several respects.  First, by restricting the
  observations to the two phases of the extrema of the binary star RV
  curve we are attempting to tailor our method to a minimum of
  observational effort as the planned observations are already very
  demanding in signal-to-noise ratio, telescope collecting area,
  and spectral resolution.  As we have shown, the CO line region in the
  K-band provides spectral features narrow enough to measure the RV
  amplitude of the companion and determine its mass from just the two
  orbital phases near the RV extrema used to produce the difference
  spectrum.  This works for a wide range of RV amplitudes (corresponding
  to different orbital inclinations and companion masses).  This type of
  direct measurement would become increasingly difficult (or less precise)
  due to line blending, if performed in difference spectra obtained
  from intermediate orbital phases.  Since they require spectra from
  several phases the approaches by \cite{simon} or \cite{hadrava} 
  should lead to a better quality of the reconstructed companion
  spectrum which, in principle, would then also allow the determination
  of the RV amplitude, provided that the various RV shifts at the
  different phases can be reconstructed as well with sufficient precision.
  However, our approach has the advantage of leading to a simple direct
  measurement that requires the observation of just two selected phases.

  Second, we apply our method to M dwarfs which, at the required
  precision level, do not have fully reliable model spectra and for
  which it is hard to find fully matching templates.  This excludes
  an approach similar to that adopted by \cite{griffin}.

  Third, the quoted methods from the literature were obviously
  developed with binary stars in mind whose components do not differ
  too much in their brightness ratio.  This is different in our work
  which focuses on studying sub-stellar companions that are much
  fainter than their host stars.  Not only do we require very high
  signal-to-noise ratios, but we also depend on a very good control
  of systematic effects such as imperfections in the wavelength
  calibration.

\section{Conclusions} 
 In this paper, we have presented a spectral differential 
technique for the detection and characterization of an 
invisible close companion to late-type stars.  
As an example an M2.5V star with a possible brown dwarf 
companion GJ1046 (\cite{kuerster}) was chosen. This system 
is very interesting as it could be the first brown dwarf 
desert object orbiting around an early-M type star. Only 
the minimum and maximum mass limits were known from 
K\"{u}rster et al.(2008). The present paper aims at developing a 
technique to determine the true companion mass.

One of the important parameters for our approach is the 
star-to-companion brightness ratio. After comparing different 
wavelength ranges, we have shown that the most suitable spectral 
region to detect the difference spectrum of the companion 
to GJ1046 is the CO line region in the K-band. 

As no high-resolution NIR spectra of GJ1046 are available in 
the data archives of the relevant observatories, we have simulated 
CRIRES observations with signal-to-noise ratio equal to 500 
in order to study the feasibility for the detection of the companion 
in the difference spectrum. We show that companion difference 
spectra with effective temperatures of 2500K, 2000K and 1500K 
would be possible to detect. 
 
One of the largest sources of systematic error comes from an 
imperfect wavelength calibration that leads to a mismatch of
the shifted stellar spectra before they are subtracted from each 
other. Our simulations show that the host star spectrum 
cancels when the mismatch is equal to 0.1 pixels or less.

As rotation of the companion star leads to broadening of the 
spectral lines that in the difference spectrum can lead to 
blending of the absorption/pseudoemission features, we have 
calculated the companion mass for different rotational velocities.
We have demonstrated that the determination of the wavelength 
shift and therefore the companion mass can be made with very 
high accuracy for slowly rotating companions. Faster rotation 
of the companion complicates its mass determination, especially 
for high companion masses. 

Our approach of subtracting the primary star spectrum
will work, if this spectrum is constant enough. Only in cases of 
extreme stellar activity may it reach its limit for our test object. We show
that the companion difference flux has larger amplitude than
the residual signal from the active star unless extreme spot 
filling factors and spot-to-photosphere temperature differences are assumed.

Knowing the true wavelength shift value we have reconstructed 
of the spectrum using singular value decomposition. From the single 
(reconstructed) spectrum of the companion, additional parameters 
can be obtained that give us important information on the companion 
object. 

\section*{Acknowledgments}
This work was supported by a DAAD stipend to N.M.
Kostogryz. M.K. Kuznetsov's work was supported by EU
PF7 Marie Curie Initial Training Networks (ITN) ROPACS
project (GA N 213646). We thank Pavlenko Ya.V. for 
his support with the WITA6 programme, F.
Rodler for helpful discussions, W. Brandner and anonymous 
referee for useful comments on our manuscript.

\label{lastpage}

\end{document}